\documentclass[aps,prb,twocolumn,showpacs,amssymb]{revtex4-1}

\usepackage[dvips]{graphicx}
\usepackage{dcolumn}
\usepackage{bm}
\def\be{\begin{equation}}
\def\en{\end{equation}}

\def\p{\partial} 

\def\tt{\tilde{\tau}} 
\newcommand{\av}[1]{\langle{#1}\rangle}
\def\gs{\gtrsim}
\def\ls{\lesssim}
\newcommand{\bi}[1]{\mbox{\boldmath$#1$}}
\def\p{\partial}

\def\bea{\begin{eqnarray}}
\def\ena{\end{eqnarray}}

\begin{document}


\title{
Self-organization  
 in  $^4$He 
 near the superfluid transition 
  in  heat flow and gravity}



\author{Syunsuke Yabunaka and Akira Onuki}
\affiliation{Department of Physics, Kyoto University, Kyoto 606-8502,
Japan}


\date{\today}

\begin{abstract} 
We investigate the nonlinear dynamics 
of $^4$He slightly below the superfluid 
transition by integrating model F equations in three dimensions. 
When a superfluid is heated from above  under gravity, 
a vortex tangle   and   sheetlike phase slips 
 both  appear  near the bottom plate. 
Then  a self-organized superfluid 
 containing high-density   vortices  
and phase slips grows upward, where 
high-amplitude second sounds are emitted from the 
self-organized to the 
 ordinary superfluid region.  
 A  phase slip 
sheet often changes into  a vortex aggregate 
and vice versa. 
The thermal resistance due to these defects 
produces a constant   temperature gradient equal to 
the gradient of the pressure-dependent 
transition temperature $T_{\lambda}(p)$. In this  self-organized  region, the temperature deviation $T-T_{\lambda}(p)$ 
consists of a negative constant 
determined by the heat flux $Q$  
and time-dependent fluctuations. 
Its time-average is calculated  
to be $155$ nK for $Q=11.2$ erg$/$cm$^2$s in good agreement 
with the experiment 
 (W.A. Moeur {\it et al.},  Phys. Rev. Lett.  {\bf 78}, 2421 (1997)). 
\end{abstract}



\maketitle


\section{Introduction}

The critical phenomena in $^4$He 
near the superfluid transition have been studied 
with exceedingly high precision of the temperature $T$ 
($\sim 10^{-9}$ deg) 
\cite{Ahlers,Ahlers80,lambda,Fmodel}.  
Such  experiments  have thus been used 
 to confirm  or predict  universal 
relations near the critical point. 
Moreover, in this system,  
  a number of  unique nonequilibrium situations 
can be found  in the nonlinear response regime.  
In particular,  the superfluid  transition  is very sensitive 
to  applied  heat flow $Q$ \cite{RMP,Onukibook}. 
 For example,    if   $T=T(z)$ is  above the transition 
temperature $T_{\lambda}$  at one end of the cell  
 and below $T_\lambda$  at the other end, 
a HeI-HeII interface  emerges  
separating  superfluid and normal fluid 
regions \cite{Erben,Bhagat,Onuki_helium,Duncan,Chui,Okuda,Dohm}. 
Here $T$  is  nearly constant in the  superfluid   
and  has a finite 
gradient in the   normal fluid, 
so the temperature gradient is almost discontinuous 
across the   interface. 
In experiments, it emerges  when $^4$He in a normal fluid 
 is cooled from a boundary  below 
$T_\lambda$ or when $^4$He in a superfluid  
  is warmed  from a  boundary  
above $T_\lambda$. This interface  
has a thickness proportional to $Q^{-1/2}$ 
at small $Q$ and is thus  a unique 
singular object.  Some authors 
were also interesred in the heat capacity 
in heat flow near 
the superfluid transition \cite{Chui}. Another 
interesting case  
is to  quench  liquid $^4$He 
from its normal to its superfluid phase 
by  a mechanical expansion \cite{Zurek,Rivers,Zurek-ex}.  
In such phase ordering with a multi-component 
order parameter, topological singularities 
(vortices in $^4$He) should be 
proliferated in an early stage after quenching 
 \cite{Zurek,Onukibook}.

The superfluid transition 
can also  be greatly influenced by 
gravity due to   the pressure 
dependence  of  
 $T_{\lambda}(p)$. 
In $^4$He near the transition,  the specific heat exhibits a sizable   
logarithmic singularity,    
but the  isothermal 
compressibility has  almost no singularity  
 \cite{Ahlers,Onukibook}, resulting in  
  nearly homogeneous density profiles 
 for not large cell length on earth.  
Thus the height-dependence 
of $T_{\lambda}(p)$ is the main 
origin of inhomogeneity 
in equilibrium $^4$He,   preventing precise 
measurements of the critical phenomena in $^4$He 
\cite{Ahlers80,lambda}.      
If  the $z$ axis 
is taken in the  upward direction  on earth, 
  $T_{\lambda}(p)$  decreases 
with the height $z$   as  
\be 
T_{\lambda}(p)= T_{\lambda{\rm bot}}( 1 + Gz),
\en 
where  $T_{\lambda{\rm bot}}$ is the transition 
temperature at the cell bottom and 
$G= |{d T_\lambda}/{d p}| \rho g/T_\lambda  \cong 0.6\times 
10^{-6}$cm$^{-1}$ at saturated vapor pressure   
with  $g$ being the gravity constant. 
The  origin of the $z$ axis is at the cell  bottom.  
We introduce the local reduced temperature  
   as 
\be
\tau \equiv 
{T}/{T_{\lambda }(p)}-1  =  
 ({T}/{T_{\lambda {\rm bot}}}-1 ) 
 - G z ,
\en 
which depends  on $z$ and represents  the distance from the  
lambda line  in the $p$-$T$ phase diagram 
Therefore, equilibrium states are  noticeably inhomogeneous 
in the temperature region 
\be
|\tau| \ls Gh,
\en   
where  $h$ is the vertical cell length.   
Remarkably, 
in the temperature range  $\tau(h)<0<\tau(0)$  
at homogeneous $T$ (in equilibrium), 
 a gravity-induced  two-phase 
coexistence is possible 
with an upper region in a superfluid 
and a lower region in a normal fluid 
\cite{Ahlers_gravity,Ginzburg}.   
The thickness $\xi_g$ and the characteristic reduced 
temperature $\tau_g$ of the gravity-induced interface 
 is determined by the balance $\tau = G\xi$, where  
 $\xi =\xi_0 |\tau|^{-2/3}$ is the correlation length 
  with $\xi_{+0}=1.4 {\rm \AA}$. 
Hence they are  given by 
\be 
\ell_g= \xi_{+0}(\xi_{+0}G)^{-2/5},
 \quad  \tau_g =(\xi_{+0}G)^{3/5},
\en  
which are about   
$10^{-9}$ and  $10^{-2}$ cm 
on earth, respectively. 
It is worth noting that application of an 
electric field $E$ or a magnetic field $H$  
gives rise to a  shift of  $T_\lambda$ 
 by an amount proportional to 
$E^2$ or $H^2$. If such a field is 
inhomogeneous, it leads to  a 
gradient of $T_\lambda$, as well as 
 the earth  gravity  \cite{Ginzburg}.

Furthermore, 
both gravity $g$ and heat flux $Q$ 
can be  crucial near the lambda line. 
Hereafter $Q$ will be measured in units of erg$/$cm$^2$s. 
If  heated  from below, 
 they both  serve to decrease 
$\tau$ with increasing the height and  
 the interface   changes over from 
the gravity-induced one to 
the heat-induced one at  $Q\sim 1$ 
with increasing $Q$. (We may determine this crossover $Q$ 
by equating the thicknesses of these interfaces.)
On the other hand,  if heated from above, 
they can compete to produce 
new  nonequilibrium  states. 
In fact, Moeur {\it et al.}\cite{Duncan_gravity,S,Duncan2007} 
realized   self-organized  superfluid in $^4$He heated from above, 
 where   the temperature gradient  and the gradient of 
  $T_\lambda(p)$ were  balanced as 
 \be 
T_\lambda \frac{d\tau}{dz}= \frac{dT}{dz} - T_\lambda G=0,
\en 
yielding a homogeneous $\tau=
-\tau_{\rm sos}(Q)$ with 
\be 
\tau_{\rm sos}(Q)  
 \cong 1.0\times 10^{-8}Q^{0.813} 
\en   
in the range  $5\ls Q\ls  60$.  
This  self-organized region 
extended over  a macroscopic region 
 in  a 7.4 mm tall, 2 cm diameter cylindrical cell. 
It is surprising that   the distance $\tau$ 
to the lambda line was held fixed  at  an extremely 
small negative constant 
over a macroscopic  region. 
Machta {\it et al.}\cite{Machta} 
 pointed out that $^4$He  in heat flow is 
   an example of self-organized criticality. 
However, $^4$He in heat flow and gravity 
stays slightly away from the lambda line, 
so it is not in a ''critical state'',   to be precise.

Notice that the balance relation (5) itself follows from the vortex 
resistance in  the Gorter-Mellink  form \cite{Donnelly},  
\be 
{dT}/{dz} = T B(T)  Q^3,
\en  
where   
$B= B_0|\tau|^{-\sigma}$ 
with $ B_0= 5\times 10^{-29}$ and $\sigma=2.23$ 
near the transition \cite{Ahlers_vortex}.  
The balance  
$B(T) Q^3 =G$ then 
yields  $\tau=- \tau_{\rm sos}^{\rm v}$ 
with
\be 
\tau_{\rm sos}^{\rm v} =
  (B_0 Q^3/G)^{1/\sigma} \cong   1.3\times 10^{-10}Q^{1.35}.
\en     
However, this  $\tau_{\rm sos}^{\rm v}$ 
 is  smaller than the experimental 
 $\tau_{\rm sos}$  in Eq.(6)  
by one or two orders of magnitude for $Q \gs 5$. 
In previous  simulations in one dimension 
\cite{Onuki_gravity2,W},
another kind of singular objects, 
phase slips,  came into play, producing 
a thermal resistance in accord with the experiment. In the 
literature, phase slips were theoretically studied 
in one  dimension 
\cite{Langer,Kramer,Ivlev} and  
have been observed in 
quasi-one-dimensional superconductors \cite{Tinkham} 
and in $^4$He in narrow apertures \cite{Sato}. 
The self-organized superfluid states 
are analogous to the resistive states 
in wire superconductors in electric field;
 however, in our case, 
phase slips appear as sheets even for 
macroscopic cell width. This is because 
the uniaxial effect of gravity  
is intensified near the lambda line. 
Similar self-organization 
should  well be realized  in $^4$He in the presence of a 
gradient of electric or magnetic field, for example, 
around the tip of  a charged needle  kept 
cooler than the surrounding liquid $^4$He.

Originally, the balance relation (5) was predicted 
 for   normal liquid $^4$He
\cite{Onuki_gravity1}. Liquid $^4$He exhibits 
strong critical fluctuations of the complex 
order parameter slightly above the lambda line, which 
 give rise to strong critical enhancement 
of the thermal conductivity $\lambda$   
\cite{Ahlers,Ahlers80,lambda,Fmodel,Onukibook}.  
For small positive $\tau$,  
experimental data of  $\lambda$  
may be fitted to the following power law \cite{lambda}, 
\be 
\lambda \cong \lambda_0 \tau^{-x_\lambda} 
 \quad  (x_\lambda \cong 0.45),
 \en   
where $\lambda_0$ is a constant. In terms of $\tau$ in Eq.(2), 
the heat conduction equation reads 
\be 
C_p \frac{\p \tau}{\p t}= \frac{\p}{\p z}\lambda\bigg[ 
 \frac{\p \tau}{\p z}+G\bigg],
\en 
where the specific heat $C_p$ is treated 
to be a constant for simplicity. 
In steady states, the temperature gradient is 
  $Q/ \lambda$ so that   
  \be 
 {d\tau}/{dz}=  ({Q}/{T_\lambda\lambda_0})
 \tau^{x_\lambda} -G  
\en   
If $\tau(h)>0$, the balance 
in Eq.(4)  gives 
$\tau=\tau_{\rm son}(Q)$ 
with  
\be 
 \tau_{\rm son}(Q) =
(T_\lambda\lambda_0 G/Q)^{1/x_\lambda} = 
3.8\times 10^{-9}Q^{-2.22}. 
\en 
This holds in the region 
 $L-z \gg \ell_n(Q)$, where  
$
\ell_n(Q)=   
\tau_{\rm son}(Q)/G
$  
is  the relaxation length.  
Weichman and Miller \cite{W} argued that 
the balance relation (5)  is attained  
 only when the correlation length 
 $\xi =\xi_0 |\tau|^{-2/3}$ 
 is shorter than $\ell_n(Q)$. 
From Eq.(4)   this local equilibrium condition 
is rewritten as $\tau_{\rm son}\gs \tau_g \sim  10^{-9}$. 
Thus we obtain the upper bound  $Q\ls 1$ 
needed  for the self-organization above the lambda line. 
 For larger $Q$,  
the system crosses across the lambda line 
and   a HeI-HeII interface appears.  
In accord with these results, 
Moeur {\it et al.}\cite{Duncan_gravity}   observed 
self-organized normal fluid states 
for $Q \ls 1$.  
Analyzing the nonlinear diffusive equation (10), 
Weichman and Miller \cite{W} also 
predicted  a new thermal 
wave propagating   only upward 
in self-organized normal 
states.  This wave was  observed
both above and below the lambda line\cite{S,Duncan2007}.

The  physical processes in the self-organized 
superfluid remain quite unclear, since in the  
experiments near the transition  
  measured was only the temperature and the simulations 
were one-dimensional. 
Realistic three-dimensional 
simulations are needed to 
investigate how  vortices and phase slips 
 come into play. 
 With this purpose,  we will present   
dynamic equations in Sec.II and  numerical results 
  in three dimensions in Sec.III.

\section{Model Equations}

 We use   the renormalized 
  model F equations \cite{Dohm,Fmodel}
 accounting for gravity \cite{Onukibook}. 
  The dynamic variables are the 
 complex order  parameter $\psi({\bi r},t)$ 
 and the entropy $s({\bi r},t)$. 
 The superfluid 
 density $\rho_s$ and the superfluid momentum density  
 ${\bi J}_s=\rho_s{\bi v}_s$ are proportional to $|\psi|^2$ and 
 ${\rm Im}(\psi^*\nabla\psi)$, respectively,  
 where ${\rm Im}(\cdots)$ 
denotes taking the imaginary part. 
The  Fourier components of $\psi$ and $s$ 
 have wave numbers smaller than a upper 
  cut-off $\Lambda$.  The fluctuations 
  with wave numbers larger than $\Lambda$ have already been 
  coarse-grained\cite{Dohm,Fmodel,Onukibook}. 
 As $\Lambda$  is decreased down to 
  the inverse correlation length 
 $\xi^{-1}$, the renormalization effect is 
 suppressed to justify use of the mean-field 
 theory at long wavelengths, where 
  the  coefficients of the model 
 are renormalized ones dependent  on $\tau$.

 In equilibrium superfluid,  the average 
 order parameter behaves as  
$\psi\propto |\tau|^{1/3}$.  Without gravity, we 
define the correlation length 
 as  $\xi= \xi_{+0}||\tau|^{-2/3}$     
 using  $\xi_{+0}=1.4 {\rm \AA}$. 
We suppose  a reference equilibrium 
  superfluid state, where  
  $\tau= -{\tt}$, 
  $\psi= \tilde{\psi}$, and $\xi=\tilde{\xi}$ 
with 
\be 
\tilde{\tau}= 2.5\times 10^{-8}, \quad 
\tilde{\xi}= 
 1.6\times 10^{-3} {\rm cm}. 
\en 
Hereafter 
 we set $\Lambda={\tilde \xi}^{-1}$ 
 and treat the fluctuations with sizes longer  
 than  $\tilde \xi$.   Those with sizes 
 shorter than $\tilde \xi$ 
 are little affected by heat flow. 
The scaled 
position ${\bi r}/{\tilde \xi}$,  
the scaled time $t/{\tilde t}$, 
the scaled correlation length 
$\xi/{\tilde \xi}$, and 
 the scaled reduced temperature 
$\epsilon/\tilde{\epsilon}$ 
will be written as ${\bi r}$, $t$, 
$\xi$, and $\epsilon$ 
in the same notation for simplicity, 
where 
\be 
\tilde{t} \cong   0.5\times 10^{-3} {\rm s} 
\en 
 is  the thermal relaxation time 
on the  scale of $\tilde\xi$. 
We introduce   the scaled reduced temperatures, 
\bea 
&&A({\bi r},t)= (T/T_{\lambda bot}-1)/\tt, 
\\   
&&\varepsilon ({\bi r},t) =\tau/\tt = A- {\tilde G}z  ,
\ena  
where ${\tilde G}=G\tilde{\xi}/\tilde{\tau}=0.04$ on earth. 
The equations for the scaled variables 
$\Psi({\bi r},t)=\psi/\tilde{\psi}$ and $A({\bi r},t)$ read \cite{Onukibook}
\bea
& &\frac{\partial}{\partial t}\Psi =i\frac{A}{a}\Psi - 
{L^*} \left [\varepsilon \xi^{-1/2}-\nabla^2+\xi^{-1} | \Psi 
|^2\right ]\Psi, \\
& &\frac{\partial}{\partial t}M=
a {\rm Im}(\Psi^*\nabla^2\Psi)  + 
\nabla\cdot {\lambda^*} \nabla A , 
\ena
where 
$M$ is the scaled entropy deviation defined by 
\be 
M= A-a^2\xi^{-1/2}|\Psi |^2/2
\en  
and  $a$ is a universal number 
of order  unity. The second term ($\propto |\Psi|^2)$ 
represents a decrease of the entropy in the ordered phase. 
We  set $a=1$  in our simulation. 
In gravity the correlation length 
does not exceeds $\ell_g$ in Eq.(4), 
so we define the local correlation length as 
\be 
\xi = \tilde{\ell}_g 
  \tanh (1/\tilde{\ell}_g |\varepsilon|^{2/3}),
\en  
where   $\tilde{\ell_g}=\ell_g/\tilde{\xi} = 1/{\tilde G}^{2/5} 
=3.62$. Then we obtain two cases of 
$\xi \cong |\varepsilon|^{-2/3}< \ell_g$ 
and   $\xi \cong \ell_g <|\varepsilon|^{-2/3}$. 
The scaled kinetic coefficients are written  as \cite{Dohm,Fmodel}
\be 
 L^*=b_\psi \xi^{0.325}, \quad  
\lambda^*
 = b_\lambda \xi^{0.675}.      
\en 
We set $ b_\psi=0.2$ and $b_\lambda=1$ in 
our simulation \cite{Onukibook}.

For $^4$He 
near the lambda line, 
the complete hydrodynamic equations 
including the dynamic equation for $\psi$  were 
presented by the Russian group many years ago \cite{Pita}.  
In the model F  \cite{Fmodel},  we 
do not treat the hydrodynamic equation 
for the mass density $\rho$ 
and the total momentum density ${\bi J}= 
\rho_n{\bi v}_n+
{\bi J}_s$  neglecting the fluctuations 
 of $\rho$ and  ${\bi J}$, 
 where ${\bi v}_n$ is the normal fluid velocity 
 and  $\rho_n=\rho-\rho_s\cong \rho$.  
Then the normal fluid velocity 
 ${\bi v}_n$ may be written as 
 \be 
 {\bi v}_n = - \rho^{-1} {\bi J}_s.
 \en  
 Therefore ${\bi v}_n$ is much smaller than  
 ${\bi v}_s= \rho_s^{-1}{\bf J}_s$ by 
 the small factor $\rho_s/\rho$. 
  This   is justified 
for slow dynamic  processes without first sounds 
near the lambda line.

In critical dynamics, 
the dynamic equations have been treated as 
Langevin equations with random source terms 
added \cite{Fmodel}. 
In this work, we neglect the random source terms in 
Eqs.(17) and (18), 
which  produce  the thermal fluctuations of 
 the order parameter $\psi$ and the entropy $s$.
They may be neglected when  the small-scale 
thermal   fluctuations 
shorter than $\xi$ 
have been coarse-grained  in the dynamic equations. 
The renormalization effect 
has already be accounted for 
in the renormalized coefficients.

\begin{figure}
\begin{center}
\includegraphics[scale=0.38]{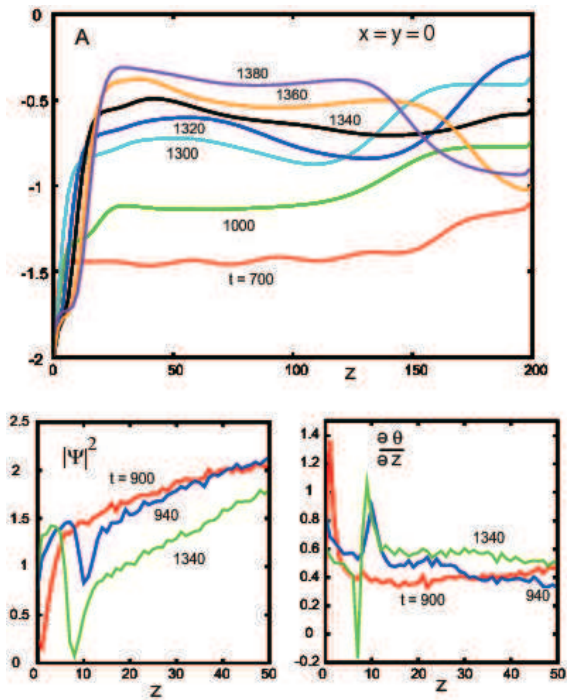}
\caption{(color online).  
Top: scaled temperature deviation $A({\bi r},t)$ 
at $x=y=0$ after application of heat flux from the top 
at $t=700, 1000,$ and $
1300+ 20n$ $(0\le n\le 4$). 
The bulk region is gradually 
heated by repeated  traversals of second sounds. 
Bottom: $|\Psi|^2 (\propto \rho_s)$ (left) 
and $\p \theta/\p z (\propto v_{sz})$ (right)  
at $t=900, 940,$ and $1340$. 
A   phase slip  
is  being created at $z\sim 10$ 
in the latter two times (see the plates 
 of Video 1). 
}
\end{center}
\end{figure}

\begin{figure}[t]
\begin{center}
\includegraphics[scale=0.44]{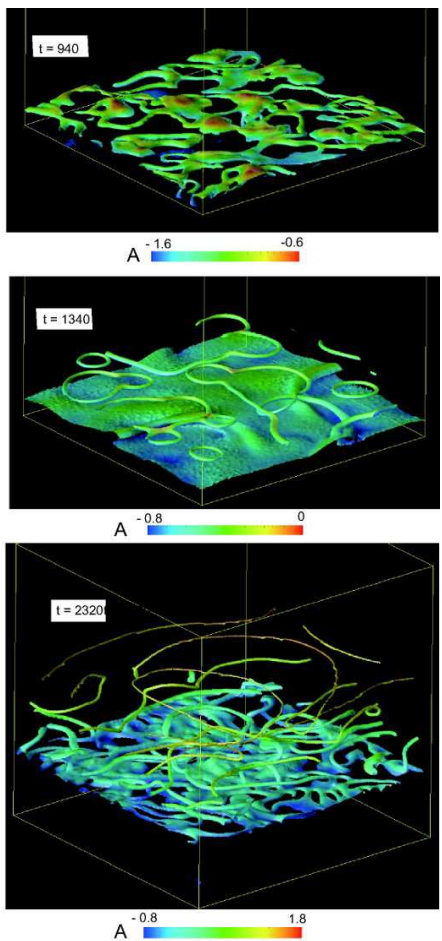}
\caption{  Surfaces  of $|\Psi|^2= 0.8$ 
 after application of heat flux  
 from the top  in a superfluid state 
in the same run as  in Fig.1. Their colors represent 
the temperature deviation $A$ with the color map below 
the plates. Top: at $t=940$ a layer in the range  
$10 \ls  z\ls 20$ is much disturbed, where a phase slip 
is going to appear. Temperature inhomogeneity 
is also apparent, for which see Fig.3.
 Middle:  at $t=1340$ a sheetlike 
 phase slip is created  near the bottom.  
 Vortices  above the sheet are expanding. 
 Bottom:  at $t=2320$ the phase slip sheet  is away from the bottom 
  and is being broken  into an aggregate 
 of vortices. 
 Video gives time evolution 
in the range $2000<t<3400$. 
}
\end{center}
\end{figure} 

\section{Simulation results}

In previous papers  \cite{Onuki_gravity2,W}, 
Eqs.(17) and (18) were integrated in one  
dimension, where phase slips 
play a major role in  self-organized superfluid 
 states, as in wire superconductors 
 in electric field \cite{Langer,Kramer,Ivlev,Tinkham}. 
  In this work, we present numerical 
 results  in three dimensions 
on a $200\times 200\times 200$ 
cubic lattice with the mesh size  equal to 
$\tilde \xi$, so the corresponding 
cell  length is $3.2$ mm. The time 
increment $\Delta t$ of integration is $0.01$ 
in units of ${\tilde t}=0.5\times 10^{-3}$s. 
The periodic boundary 
conditions are imposed in the $xy$ plane.  
At $z=0$ and $200$,  
we set $\Psi=0$ and assume  homogeneity 
of the  temperature and the heat 
flux in the $xy$ plane, supposing  metallic plates. 
We  prepared  an equilibrium 
 superfluid state with  $A=-2$ for $t<0$. 
 We then applied   a constant heat flux at   $z=h$,   
but we kept   the bottom temperature 
at $A=-2$ ($\tau= -2{\tilde \tau}$)  
at $z=0$ for $t>0$.  The  scaled heat flux 
$\hat{Q}=\lambda^* (\p A/\p z)$ 
at  the top is equal to 1. This  corresponds   
to a heat flux of $Q=11.2$, which 
 is the value of 
$\lambda T_\lambda \tau/\xi$ at $\tau=\tilde{\tau}$.
In the following we will explain our numerical results.

In Fig.1, we display profiles of 
the scaled reduced temperature $A$ 
vs $z$ (with  $x=y=0$)  at seven  times 
$t=700, 1000$, and $ 1260+ 40n$ $(0\le n\le 4$), 
where traversals of second sounds 
cause gradual heating of the  bulk superfluid 
\cite{Ahlers_gravity}. 
Since the bottom temperature is fixed, 
a transition layer supporting a temperature gradient 
grows  near the bottom,  where   
a heat flux of order 0.1 
is from the liquid to  the bottom plate. 
This  heat output is  $10\%$ of  the heat input from 
the top.  In the lower plates, 
we show  $|\Psi|^2$  and $\p \theta/\p z$ 
 near the bottom at three times to illustrate 
 formation of a  phase slip, where $\theta$ is the phase 
 of $\Psi=|\Psi|e^{i\theta}$. We confirmed that 
 this phase slip formation in a narrow region 
 occurs  even if the mesh size is reduced to $0.2{\tilde \xi}$.  
Here $\p \theta/\p z$  is equal to 
 the superfluid velocity 
 in the $z$ axis in units of $\hbar/m\bar{\xi}$, 
 where $m$ is the $^4$He mass. 
Note that the complex order 
parameter behaves as 
$\psi\propto e^{ikz}$  with $v_s=\hbar k/m$ 
in current-carrying   superfluids    
and the bulk region is unstable 
  for $k$ larger than 
  $1/\sqrt{3}{\xi}$ from linear stability 
  analysis\cite{Kra}.  In our case,  
  $\rho_s$ decreases towards zero 
  and  $v_s$ exceeds   the critical value near the bottom, 
 resulting in  defect formation.

 In Video 1, 
we display the  surfaces of  
  $|\Psi|^2= 0.8$ near the bottom, which illustrates 
   growth of vortices and phase slips. 
  At  $t=940$  disturbances of $\Psi$ are enhanced  
 in  a layer in the range  
$10 \ls  z\ls 20$. See  the lower plates in Fig.1. 
We can also see considerable  
inhomogeneity of the temperature. 
At $t=1340$,  vortices 
   are expanding and  a 
 sheetlike  phase slip is at its birth. 
In our case  under gravity, $|\Psi|^2$ decreases near the bottom. 
We observe that  undulations on  the  sheet 
grow and  detach   to form  vortices.     
At $t=2320$, the sheet is ramified into 
 a vortex  aggregate. 
Afterwards, sheetlike phase slips successively 
appear  forming an expanding 
 self-organized superfluid. 
As well as vortex proliferation, 
 vortex aggregates 
sometimes fuse  into sheetlike 
phase slips. These temporal  changeovers  
occur   on  the system scale 
 with large-amplitude 
second sounds. No clear boundary can be seen between 
the self-organized and ordinary superfluid regions, 
but high-density vortices are undergoing  
large-scale temporal fluctuations in the 
transition region. 

\begin{figure}
\begin{center}
\includegraphics[scale=0.4]{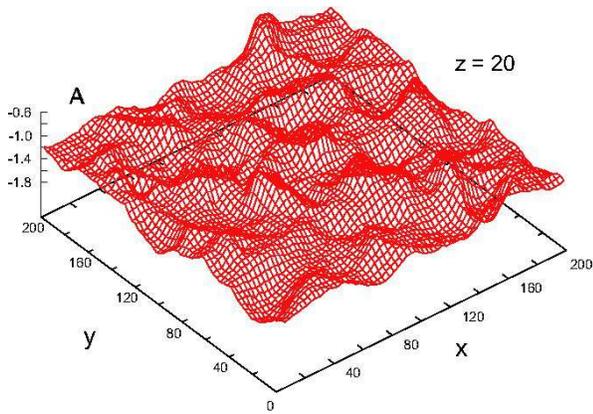}
\caption{(color online). 
Scaled temperature deviation $A({\bi r},t)$ 
in the $x$-$y$ plane 
at  $z=20$ taken at $t=940$ in the same run 
as in Fig.1 and Video 1, corresponding to 
 the top plate in Video 1. 
A phase slip is being created 
with considerable temperature inhomogeneity, 
where  $\av{A}= -1.22$  
 and $\av{(A-\av{A})^2}= 0.156^2$ with $\av{\cdots}$ 
denoting  the areal average.  
}
\end{center}
\end{figure}

\begin{figure}[t]
\begin{center}
\includegraphics[scale=0.4]{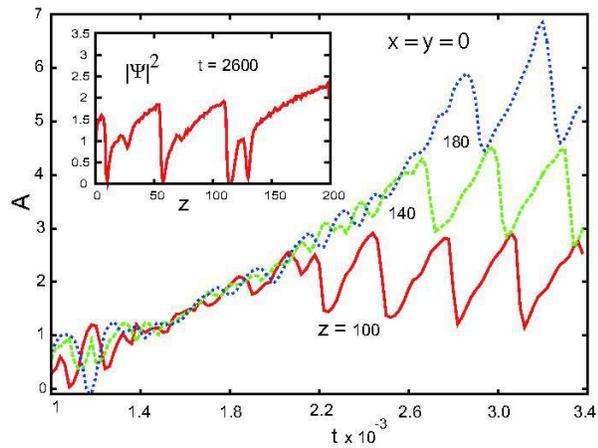}
\caption{(color online). Scaled temperature deviation $A({\bi r},t)$ 
at $z=100$, $140$, and $180$ 
with  $x=y=0$ after application of heat flux from the top 
 in the time range  $1000<t<3400$  in the same run 
as in Figs.1 and 2. Inset: $|\Psi|^2$ vs $z$ at $t=2600$,  
which is zero at defect points. 
At this time,   self-organized and ordinary 
superfluid regions are separated  at $z\sim 130$.  }
\end{center}
\end{figure}

In Fig.3, the inhomogeneity of 
$A({\bi r},t)$ is shown 
in the  $x$-$y$ plane 
at  $z=20$ taken at $t=940$. 
It originates from  
the relation (16) for the entropy 
and the temperature. That is, if $\rho_s$  
increases in some regions, 
the temperature there tends to increase 
if the entropy change is slower.

In Fig.4,   time evolution  of $A$ is shown 
at three heights  with $x=y=0$ in a longer 
time range $10^3<t<3.4\times 10^3$ 
in the same run producing Figs.1 and 2. 
For $t<2000$ the three points are 
in an ordinary superfluid  
and are nearly uniformly heated. Subsequent zigzag 
temperature changes are due to the passage of platelike 
defect aggregates. 
A similar plot was given by Moeur {\it et al.}
 \cite{Duncan_gravity}, 
where the temperatures at three thermometers 
attached to the side wall became flat during the passage 
of a self-organized region. 
In our  simulation, a train of defects eventually 
arrived  at the top and a normal fluid 
with  a steep temperature gradient emerged   from 
the top.

\begin{figure}[t]
\begin{center}
\includegraphics[scale=0.41]{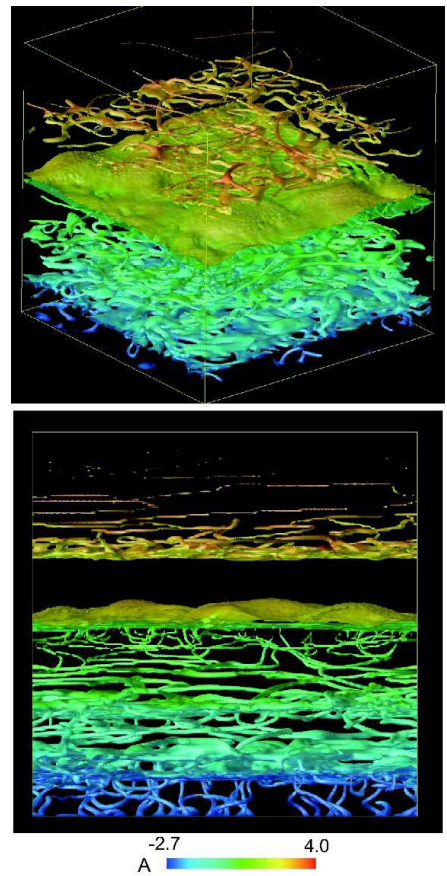}
\caption{ Surfaces  of $|\Psi|^2= 1.2$ 
 viewed from two angles 
 in a dynamical steady state 
 at $t=3840$.  
 Their colors represent 
the temperature deviation $A$ with the color map below 
the plates. A sheetlike phase slip can be seen at the middle, 
while the other defects are aggregates of vortices. 
 }
\end{center}
\end{figure}

\begin{figure}[t]
\begin{center}
\includegraphics[scale=0.4]{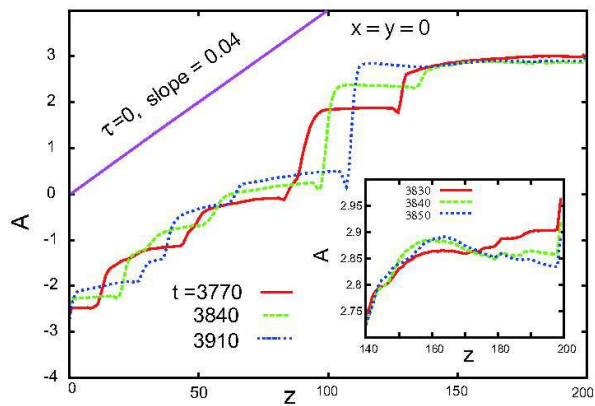}
\caption{(color online). Scaled temperature deviation $A({\bi r},t)$ 
vs $z$ ($x=y=0$)  at  $t= 3770, 3840$, and 3910 
in a dynamical steady state    in the same run as in Fig.4. 
Heat fluxes 
at the top and bottom are fixed at a common value. 
Significant  defect movements    can be seen. 
Line of $\tau=0$ is also shown. 
Inset: $A$ vs $z$ at $t=3830$, 3840, and 3850 
in the  ordinary superfluid 
region, indicating   standing second sound waves. 
}
\end{center}
\end{figure}

\begin{figure}[t]
\begin{center}
\includegraphics[scale=0.4]{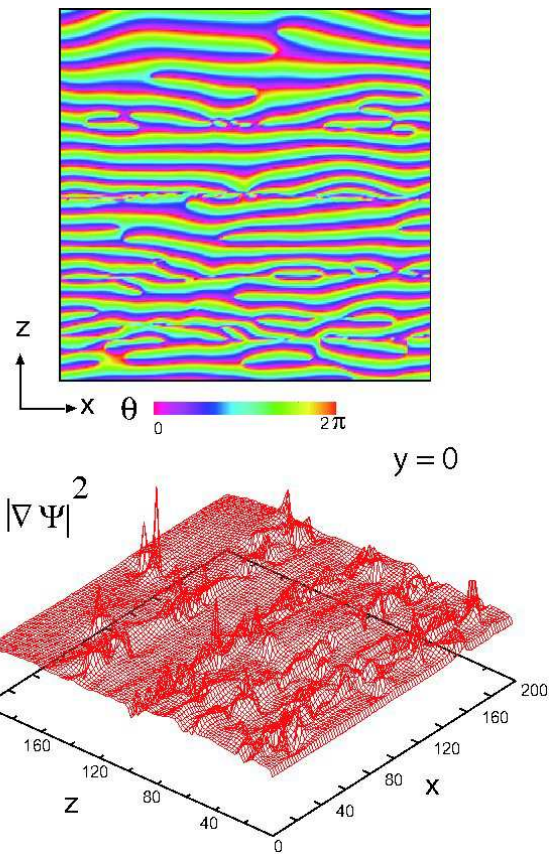}
\caption{(color online). Phase $\theta({\bi r},t)$ 
 of $\Phi({\bi r},t)$ (upper plate) 
 in the range $0\le \theta<2\pi$ 
 according to the color bar 
and $|\nabla\Phi|^2$ (lower plate) 
in the $x$-$z$ plane at $y=0$  
 taken  at  $t=  3840$, where  use is made of 
 the same data as in Fig.5. 
 }
\end{center}
\end{figure}

In Fig.4, the advancing 
speed of the self-organized region 
is about  $0.1$ (in units of ${\tilde \xi}/{\tilde t})$ 
or $3$ cm$/$s,  while it was  
$2\times 10^{-3}$cm$/$s 
in one of the experimental  data  
(see Fig.2 of Ref.19).
Note that  the speed of this  phase change 
is determined by the rate of heat input to the system. 
To  check   this aspect,   
we fixed  the bottom heat flux 
at  the top value 1 and resumed  the integration 
using the data at $t=2500$ in the  run producing   
Figs.1-4. 
Then we could realize a dynamical steady  state 
 with a self-organized superfluid in a lower part   
and an ordinary superfluid  in an upper part. 
Hereafter the times  are 
2500 plus those after the restart.

In Fig.5, we display  
snapshots of the surfaces of $|\Psi|^2=1.2$ 
at $t=3840$ from above and side. 
The defect structure here 
is composed of   a single 
sheetlike  phase slip in the middle 
and platelike vortex aggregates. 
The vortices are highly  connected 
one another forming a network. 
Some vortices   start from 
the phase slip sheet or from  the bottom 
wall. In this configuration,  defects are absent 
 in a layer region 
above the phase slip plane, but above 
this layer  a vortex aggregate 
appears to form  an  interface between the 
self-organized and ordinary superfluid regions.

In Fig.6, we present  the scaled 
temperature deviation $A({\bi r},t)$ 
 vs $z$ with $x=y=0$    at three  
consecutive  times in the dynamical steady state. It  shows    
considerable displacive motions  
  of the defect structure 
 on  a   time scale  of 70. 
The temperature variations are   stepwise at 
a phase slip and are more gradual   across vortex aggregates. 
In accord with the balance relation (5), 
the zig-zag curves of $A$ are separated 
 from the line of $\tau=0$ by 
2.5, which  is our numerical  result  of 
$\tau_{\rm sos}/{\tilde \tau}$. 
In agreement with this result,  the  experimental 
formula  (6) yields 
\be 
\tau_{\rm sos}/{\tilde \tau}=11.2^{0.813}/2.5=2.85.
\en  
Note that  phase slips in   the  one-dimensional 
simulation \cite{Onuki_gravity2} already 
gave $\tau_{\rm sos}$ 
 in agreement with  the experiment\cite{Duncan_gravity}. 
In the inset, we demonstrate 
 the presence of second sounds 
  oscillating on 
 a time scale  of 10 (smaller than those in Fig.1) in 
 the upper ordinary superfluid region.

In Fig.7, we show the phase $\theta $ 
and the gradient amplitude $|\nabla \Psi|^2$ 
in the $x$-$z$ plane at $y=0$  
using the same data as in Fig.5 at  $t=  3840$. 
These are  cross-sectional profiles  illustrating  the 
complex defect structure.  Around the defects,  $\theta$  
 varies   steeply in the upper plate, while 
 peaks and ridges 
 appear in the 
lower plate. Here    we have 
 $|\nabla \Psi|^2\cong  |\Psi|^2 
|\nabla \theta|^2 \propto \rho_s v_s^2$ around the defects.

\section{Summary and remarks }

In summary, we have investigated 
 the self-organized superfluid state  
 on earth  under a fixed downward heat flux $Q=11.2$ 
erg$/$cm$^2$s from the top in a cubic 
cell of $3.2$ mm length. 
The periodic boundary condition 
has been imposed  in the horizontal 
directions.

First, at a fixed bottom temperature,  
we have examined how   
vortices and phase slips appear 
near the bottom and how a  
self-organized region expands upward. 
Second, by setting  the bottom heat flux 
equal to the top heat flux, 
we have realized  a dynamical 
steady state in which the 
self-organized and ordinary superfluids 
coexist. The space-time fluctuations 
are highly enhanced, where   a sheetlike 
phase slip often changes into 
 a vortex aggregate and 
vice verse.  In fact, 
in Fig.5, we can see  only a single  
sheetlike phase slip, while  
numerous vortex lines form platelike  networks.  
Our numerical value of $\tau$ in Eq.(2) 
(the distance from the lambda line) in 
 self-organized superfluid states  
is in good agreement with 
the experimental $\tau_{\rm sos}$ in Eq.(6). 

We   estimate  the temperature drop 
at a phase slip   as  
\be 
(\Delta T)_p= 
Q\ell_{\rm p}/\lambda(\tau_{\rm sos}),
\en  
where  $\ell_{\rm p}$ is the thickness of a  
phase slip ($\sim \xi$) and $\lambda(\tau_{\rm sos})$ 
is the thermal conductivity in Eq.(9) 
at $\tau=\tau_{\rm sos}$.   
If the defects consist of phase slips only, 
their density $n_{\rm p} (<\ell_{\rm p}^{-1})$ per unit length 
satisfies $n_{\rm p}(\Delta T)_{\rm p}= T_\lambda G$ 
in  self-organized superfluid states. Use of Eqs.(9), (12), 
and (24) yields 
\be 
n_{\rm p}\ell_{\rm p}
 = GT_\lambda \lambda(\tau_{\rm sos}) /Q
= (\tau_{\rm son}/\tau_{\rm sos})^{x_\lambda}. 
\en  
This estimation is consistent with  the results in the 
one-dimensional simulations \cite{Onuki_gravity2,W}. 
Here the required inequality  
 $n_{\rm p}<\ell_{\rm p}^{-1}$ 
 holds for $Q>1$. In accord with this,  
  self-organized superfluid states were 
observed only for $Q>1$ in the experiment \cite{Duncan_gravity}. 
 Notice that  the  vortex aggregates in Fig.5 
also give rise to temperature drops 
comparable to that of a phase slip as in Fig.6.

A self-organized superfluid expands upward 
if the heat input from the top is larger than 
the heat output from the bottom. 
After its arrival at the top, 
 a normal fluid begins to expand downward 
 (not shown in this paper). 
Therefore,  there can also be 
coexistence of a self-organized superfluid  
and a  normal fluid. 
This coexistence  was realized in 
 our previous simulation \cite{Onuki_gravity2}   
 by cooling   a normal fluid 
 at  the bottom below the transition. 
In such cooling, a self-organized superfluid region 
grows upward to reach the top and 
then an ordinary superfluid region 
appears to expand downward.    On heating or cooling from a 
boundary wall without gravity or at large $Q$, 
a HeI-HeII interface 
appears and  its movement leads to 
the phase transformation between normal fluid and superfluid.

Phase slips 
are well-known singular 
entities observed in quasi-one-dimensional 
 superconductors \cite{Tinkham}
  and $^4$He \cite{Sato}. 
However,  under an external potential, they can emerge 
as sheets in three-dimensional 
geometries. 
The  role  of such a potential 
becomes relevant 
on approaching the criticality. 
In the case of gravity, 
this crossover occurs extremely close to 
the lambda line.  Self-organized superfluidity 
may be realized more easily  in   an inhomogeneous,   
intense electric or magnetic field, 
 as  stated  in Sec.I,  
or under high-speed rotation.

In the literature\cite{Zurek,Rivers,Zurek-ex}, 
 vortex generation after  quenching of  $^4$He 
was proposed but it has  not 
been detected in subsequent  experiments, where  
crossing of the lambda line 
by a mechanical    expansion  
is supposed.  A similar  quenching method 
(using the piston effect) 
was used to induce phase separation 
in fluids near the gas-liquid critical point 
\cite{Be}.  In  a decompression experiment 
near the superfluid transition, however,  
we should examine   how 
the  phase transition is induced 
by  sounds.    As recently reported \cite{Miura}, 
  first sounds   emitted from 
a  boundary wall  
propagate throughout the cell 
with a stepwise pressure decrease 
 at the wave front. 
They are    multiply  reflected 
at  the boundary walls 
until  dissipated. If the sound speed is $10^4$cm$/$s 
and the cell length is 1 cm,    
the  acoustic traversal time 
is $10^{-4}$s 
and is much longer than the  
relaxation time of the 
order parameter 
except extremely close to 
the lambda line.  
Second sounds are also emitted from the boundary walls  
with appearance of superfluid.   
Further investigation of  these processes   
is thus needed.

\begin{acknowledgments}
One of the authors (A.O.) 
thanks  R. Duncan for 
stimulating discussions in an early stage of this work. 
This work was supported by Grants-in-Aid 
for   the Global COE program 
``The Next Generation of Physics, Spun from Universality and Emergence" 
of Kyoto University 
 from the Ministry of Education, 
Culture, Sports, Science and Technology of Japan. 
\end{acknowledgments}


\end{document}